THIS IS A LaTeX FILE.  RUN TWICE TO GET CROSS REFERENCES RIGHT.
ALL MACROS ARE INCLUDED.
\documentstyle[11pt]{article}

\newcommand{\be}{\begin{equation}}
\newcommand{\ee}{\end{equation}}
\newcommand{\bea}{\begin{eqnarray}}
\newcommand{\eea}{\end{eqnarray}}
\newcommand{\beann}{\begin{eqnarray*}}
\newcommand{\eeann}{\end{eqnarray*}}
\newcommand{\beasn}{\begin{sneqnarray}}
\newcommand{\eeasn}{\end{sneqnarray}}
\newcommand{\bref}[1]{(\ref{#1})}
\newcommand{\eps}{\epsilon}

\newcommand{\NPB}[3]{{\sl Nucl. Phys.} {\bf B#1} (19#2)  {#3}}

\newcommand{\PLB}[3]{{\sl Phys. Lett.} {\bf #1B} (19#2)  {#3}}

\newcommand{\IJA}[3]{{\sl Int. J. Mod. Phys.} {\bf A#1} (19#2) {#3}}
\newcommand{\CMP}[3]{{\sl Commun. Math. Phys.} {\bf #1} (19#2) {#3}}

\catcode`@=11
\def\dif{{\rm d}}
\def\deriv{\@ifnextchar[{\@deriv}{\@deriv[]}}
   \def\@deriv[#1]#2#3{\mathchoice%
{{\dif^{#1}#2\over\dif{#3}^{#1}}}{{\dif^{#1}#2/\dif{#3}^{#1}}}%
{{\dif^{#1}#2\over\dif{#3}^{#1}}}{{\dif^{#1}#2/\dif{#3}^{#1}}}}

\def\presup#1{{}^{#1}\kern-.15em\relax}      %pre-superscript
\def\presub#1{{}_{#1}\kern-.12em\relax}      %pre-subscript

\def\secteqno{\@addtoreset{equation}{section}%
\def\theequation{\thesection.\arabic{equation}}}
\def\endsecteqno{\def\theequation{\@ifundefined{chapter}%
{\arabic{equation}}{\thechapter.\arabic{equation}}}}
\newcounter{subequation}
\def\thesubequation{\alph{subequation}}
\def\sneqnarray{\stepcounter{equation}\let\@currentlabel=\theequation
\setcounter{subequation}{1}
\def\@eqnnum{{\rm (\theequation\thesubequation)}}
\global\@eqcnt\z@\tabskip\@centering\let\\=\@eqncr\let\@@eqncr=\@@sneqncr
$$\halign to \displaywidth\bgroup\@eqnsel\hskip\@centering
 $\displaystyle\tabskip\z@{##}$&\global\@eqcnt\@ne
 \hskip 2\arraycolsep \hfil${##}$\hfil
 &\global\@eqcnt\tw@ \hskip 2\arraycolsep $\displaystyle\tabskip\z@{##}$\hfil
  \tabskip\@centering&\llap{##}\tabskip\z@\cr}
\def\endsneqnarray{\@@sneqncr\egroup $$\global\@ignoretrue}
\def\@@sneqncr{\let\@tempa\relax
   \ifcase\@eqcnt \def\@tempa{& & &}\or \def\@tempa{& &}
   \else \def\@tempa{&}\fi
     \@tempa \if@eqnsw\@eqnnum\stepcounter{subequation}\fi
     \global\@eqnswtrue\global\@eqcnt\z@\cr}
\def\nobiblabels{\def\@lbibitem[##1]##2{\@bibitem{##2}}}
\catcode`@=12
\def\dddot#1{\hbox{$\mathop{#1}\limits^{\ldots}$}}

\def\W{$\cal W$}

\secteqno

\title{\bf Zero-curvature condition in two dimensions.\\
           Relativistic particle models and \\
           finite \W-transformations}

\author{{\sc J. Gomis},$^\dagger$%
        \thanks{On leave of absence from Dept.\ d'Estructura i
                Constituents de la Mat\`eria, U. Barcelona.}\,
        {\sc J. Herrero},$^\sharp$
        {\sc K. Kamimura}$^\flat$
        {\sc and J. Roca}$^\sharp$\\
        \llap{$^\dagger$}%
        \small{\it{Theory Group, Department of Physics}}\\
        \small{\it{The Univ.\ of Texas at Austin}}\\
        \small{\it{RLM\,5208, Austin, TEXAS}}\\
        \llap{$^\sharp$}%
        \small{\it{Departament d'Estructura i Constituents
               de la Mat\`eria}}\\
        \small{\it{Universitat de Barcelona}}\\
        \small{\it{Diagonal, 647}}\\
        \small{\it{E-08028 BARCELONA}}\\
        \llap{$^\flat$}%
        \small{\it{Department of Physics, Toho University}}\\
        \small{\it{Funabashi}}\\
        \small{\it{274 Japan}}\\
        {\it e-mails:} \small{GOMIS@UTAPHY, HERRERO@EBUBECM1,}\\
                       \small{KAMIMURA@JPNYITP, ROCA@EBUBECM1}}

\date{}

\begin{document}

\maketitle

\thispagestyle{empty}

\begin{abstract}

A relation between an $Sp(2M)$ gauge particle model and the zero-curvature
condition in a two-dimensional gauge theory is presented. For the $Sp(4)$
case we construct finite \W-transformations.

\end{abstract}

%\newpage
%\def\litAbstract{Resum}
%\begin{abstract}
%
%Presentem una relaci\'o entre un model gauge $Sp(2M)$ de particula i la
%condici\'o
%de curvatura nu\ll a en una teoria gauge en dues dimensions. Per al cas
%de $Sp(4)$ construim transformacions \W\ finites.
%
%\end{abstract}

\vfill\hfill
\vbox{
\hfill March 1993\null\par
\hfill UTTG-04-93\null\par
\hfill UB-ECM-PF 93/6\null\par
\hfill TOHO-FP-9344}\null

\clearpage

\section{Introduction}

\hspace{\parindent}%
In the last few years a lot of attention has been devoted to the
study of \W-algebras \cite {Z}. For recent update reviews see \cite {R},
\cite {B} where extensive lists of references can be found.

  An interesting way to construct classical \W-algebras is by the
zero-curvature method \cite {P}, \cite {DS}, \cite {BFK},\cite {D} \cite {BG}.
If one constraints an $A_z$ gauge potential the
residual gauge transformations can be obtained as a zero-curvature
condition $F_{z\bar z}=0$.
This zero-curvature condition is the integrability condition
of a linear system of partial differential equations.
As we will see this
system can be related to the transformation properties and equations
of motion of  matter coupled to the gauge fields.

In this letter we consider a relativistic model of $M$ particles with
an $Sp(2M)$ gauge group, the matter variables being the coordinates and
momenta of the particles and  the gauge variables being the Lagrange
multipliers.
We find that under some formal identifications between $2d$ gauge
theories and $1d$ particle models, the equations of motion and
transformation properties of the matter variables can be written as a
system of partial differential equations whose integrability condition
is precisely the zero-curvature condition $F_{z\bar z}=0$.
This condition is equivalent to the transformation properties of the
Lagrange multipliers.
These relations continue to hold when we fix the gauge partially.
This fact explains why a model of relativistic particles
exhibits, after a partial gauge-fixing, invariance under non-linear
\W-symmetry tranformations. In a sense, it can be understood as a
coupling of matter to (world-line) \W-gravity.

The particle model is also useful for the construction of finite
\W-transformations.
Finite transformations are necessary in order to understand completely
the \W-geometry \cite{BFK}, \cite{BG}, \cite{wgeo1}, \cite{G},
\cite{wgeo2}, \cite{wgeo3}, \cite{wgeo4}, \cite{wgeo5}, \cite{wgeo6}.
The strategy is the following: we first construct the finite linear
transformations of the $Sp(2M)$ model and then, by a partial gauge-fixing
at the finite level, we find residual finite \W-transformations. In this
way one avoids the direct integration of non-linear infinitesimal
\W-transformations. We will explicitly construct in this paper
finite \W-transformations obtained from the $Sp(4)$ gauge group.

\section{$Sp(2M)$ model}

\hspace{\parindent}%
Let us consider a reparametrization-invariant model of $M$ relativistic
particles with an $Sp(2M)$ gauge group living in a flat non-euclidean
$d$-dimensional
 space-time. The dimension $d$ is large enough so the
constraints do not trivialize the model.
The canonical action is given by
\be
S=\int\dif t\left(p_i\dot x_i-\lambda_{A_{ij}}\phi_{A_{ij}}\right),
\quad\quad i,j=1, \ldots ,M,\quad A=1,2,3.
\label{sp can act}
\ee
The variable $x_i^\mu(t)$ is the world-line coordinate of the
$i$-th particle and
$p_i^\mu$ is its corresponding momentum. The Lagrange multipliers
$\lambda_{A_{ij}}$ implement the constraints $\phi_{A_{ij}}=0$ and
satisfy

$$
\lambda_{1_{ji}} = \lambda_{1_{ij}},\quad\quad
\lambda_{3_{ji}} = \lambda_{3_{ij}}.
$$
The explicit form of $\phi_{A_{ij}}$ is
\be
\phi_{1_{ij}}=\frac 12 p_ip_j,\quad\quad
\phi_{2_{ij}}=p_ix_j\quad\quad {\rm and} \quad\quad
\phi_{3_{ij}}=\frac 12 x_ix_j.
\ee

These $~2M^2+M~$ constraints close under Poisson bracket giving
a realization of the $Sp(2M)$ algebra.

It is useful to introduce a matrix notation for the coordinates and
momenta of the particles
\be
R=\left(\begin{array}{c}
        r\\
        p
        \end{array}\right),\quad\quad
\bar R=\left(p^\top,\;-r^\top\right)
\ee
with
$$
r=\left(\begin{array}{c}
         x_1\\
         \vdots\\
         x_M
         \end{array}\right),\quad\quad p=\left(\begin{array}{c}
                                            p_1\\
                                            \vdots\\
                                            p_M
                                            \end{array}\right).
$$
We can put the Lagrange multipliers in a $2M \times 2M$ symplectic
matrix

\be
\Lambda=\left(\begin{array}{cc}
              B&A\\
             -C&-B^\top
              \end{array}\right)
\ee
where the components of the $M \times M$ matrices $ A,B ,C$ are the
Lagrange multipliers $\lambda_{1_{ij}},~\lambda_{2_{ij}},~
 \lambda_{3_{ij}}$ respectively.

The canonical action \bref{sp can act} can be written in a matrix
form as
\be
S=\int\dif t~\frac12\left(\bar R\dot R-\bar R\Lambda R\right).
\ee

In this formulation the gauge invariance of the action is  given
by ordinary Yang-Mills type transformations
\footnote{For a previous discussion of geometrical models and Yang-Mills
gauge theories see \cite{K}.}
 with the gauge group $Sp(2M)$:
\be
\delta R=\beta R,
\ee
\be
\delta\Lambda=\dot\beta-[\Lambda,\beta],
\ee
where $\beta$ is the $2M \times 2M$ matrix of gauge parameters
\be
\beta=\left(\begin{array}{cc}
              B_\beta&A_\beta\\
             -C_\beta&-B^\top_\beta
              \end{array}\right)
\ee
and $A_{\beta},B_{\beta},C_{\beta}$ are the $M \times M$
matrices gauge parameters associated to
the constraints $\phi_{1_{ij}},\phi_{2_{ij}},\phi_{3_{ij}}$.

The equations of motion of the matter fields are
\be
\dot R-\Lambda R=0.
\ee

If we make the following identifications
\be
\delta \rightarrow \bar{\partial}
,\quad \beta \rightarrow A_{\bar{z}}
,\quad \Lambda \rightarrow A_z
,\quad \frac {\dif}{\dif t} \rightarrow \partial,
\ee
the equations of motion (2.9) and transformation properties of
the matter fields (2.6) can be written as
\be
\nonumber
(\bar{\partial}-A_{\bar{z}})R=0,~~~~
\\\nonumber
(\partial-A_z)R=0.
\ee
This linear system of partial differential equations has an integrability
condition ~~~~$F_{z\bar{z}}=0 $, which is equivalent
to the transformation law of the Lagrange multipliers (2.7)
under  the identifications (2.10). These relations will
continue to hold when we fix the gauge partially.
The above discussion explains why a relativistic particle model
becomes after partial gauge-fixing a model of matter with a
non-linear \W-symmetry.

As we will see below it is useful to express the model in terms of
lagrangian variables in order to construct the finite transformations
of the model.
If we write the momenta $p$  in terms of the lagrangian variables
\be
p=A^{-1}(\dot r-Br)\equiv K,
\ee
 the action is now rewritten as
\be
S=\int\dif t~\frac12\left(K^\top AK-r^\top Cr\right).
\label {al}
\ee
The gauge transformations are
\be
\delta r=A_\beta K+B_\beta r,~~~~
\delta\Lambda=\dot\beta-[\Lambda,\beta].
\label {open}
\ee

A characteristic feature of these lagrangian transformations is that
the algebra is open, except for $Sp(2)$,
$$
\nonumber
[\delta_1,\delta_2]\;r=\delta_{\beta^*} r+
(A_{\beta_2}A^{-1}A_{\beta_1}-A_{\beta_1}A^{-1}A_{\beta_2})[L]_r,
$$
\be
\\\nonumber
[\delta_1,\delta_2]\;\Lambda=\delta_{\beta^*}\Lambda,
\ee
where $\beta^*=[\beta_2,\beta_1]$ and $[L]_r$ are the Euler-Lagrange
equations of motion of $r$.
There are two reasons for the appearance of an open algebra:
1) the transformations of the momenta at the
lagrangian and hamiltonian level do not generally coincide,
2) there are more than one first-class constraints quadratic in the
momenta.

In order to close the gauge algebra we introduce $M$ auxiliary vectors
$(F_1,...F_M)$ and modify the transformation law of the coordinates $r$ as
\be
\delta r=A_\beta(K+F)+B_\beta r.
\label{mat tra}
\ee

The transformation of $F$ is determined by the condition that $K+F$
transforms as $p$ in the hamiltonian formalism. Explicitly we  get
\be
\delta F=-A^{-1}\left[A_\beta\partial_t(K+F)+
A_\beta B^\top(K+F)+(\delta A-B_\beta A)F+A_\beta Cr\right],
\label{aux tra}
\ee
while the transformation of $\Lambda$ remains unchanged
\be
\delta\Lambda=\dot\beta-[\Lambda,\beta].
\label{mul tra}
\ee
 The new algebra closes off-shell.

The  invariant action under the modified gauge transformations
is
\be
S=\int\dif t ~\frac12 \left(K^\top AK-r^\top Cr-F^\top AF\right).
\ee
The redundancy of the auxiliary variables $~F~$ is guaranteed by the
action itself.
This action is also invariant under ordinary diffeomorphisms (Diff).
\be
\nonumber
\delta_Dr=\eps\dot r,~~~~
\\\nonumber
\delta_DF=\eps\dot F,~~~~
\\\nonumber
\delta_D\lambda_j=\partial_t(\eps\lambda_j).
\ee

Diffeomorphisms are not an independent symmetry on-shell.
In fact they can be obtained from $\beta$ transformations by setting
$~\beta_{A_{ij}}=\epsilon\lambda_{A_{ij}}~$
up to some terms vanishing on the equations of motion.
Diffeomorphisms and $\beta$ transformations form a closed algebra
\be
[\delta_\beta,\delta_D]=\delta_{\beta'},\quad\quad{\rm with}\quad
\beta'=\eps\beta_.
\ee

As we want to obtain \W-transformations after a partial gauge-fixing
and these transformations contain ordinary Diff it is natural to consider
Diff and $(M(2M+1)-1)$ of $\beta $ transformations as the independent
symmetry transformations.

\section{Finite transformations of $Sp(2M)$ model}

\hspace{\parindent}%
Now we want to find the finite linear transformations of this model.

Finite trans\-form\-ations \footnote{
For a recent discussion on finite gauge transformations see \cite{GPR}.}
can be obtained by exponentiating the
infinitesimal ones as
$
{X^i}'=\exp\{\theta^\alpha\Gamma_\alpha\}X^i,
$
where the generators $\Gamma_\alpha=R^i_\alpha\frac{\partial}
{\partial X^i}$ satisfy
$
[\Gamma_\alpha,\Gamma_\beta]=f^\gamma_{\alpha\beta}\Gamma_\gamma
$
and $ X^i$ represents any of the variables.
The coefficients $f^\gamma_{\alpha\beta}$ are the structure functions of
the $Sp(2M)$ gauge algebra.

It is useful to perform the integration using the matrix notation.
The explicit form of the finite gauge transformations is considered
in the following four sets of transformations. Any finite transformation
may be obtained by the composition of them.
\vskip 3mm

$\bullet$ Transformations generated by $A_\beta$:
$$
A'=A+\{\dot A_\beta-A_\beta B^\top-BA_\beta\}+A_\beta
CA_\beta,
$$
$$ B'=B-A_\beta C,\quad\quad\quad C'=C,
$$
$$r'=r+A_\beta(K+F),
$$
\be
F'={A'}^{-1}\left[AF-A_\beta\{\partial_t(K+F)+B^\top(K+F)+Cr\}\right].
\ee
\vskip 3mm

$\bullet$ Transformations generated by $B_\beta$:
$$
A'=e^{B_\beta}Ae^{B^\top_\beta},~~~~
B'=e^{B_\beta}(B-\partial_t)e^{-B_\beta},~~~~
C'=e^{-B^\top_\beta}Ce^{-B_\beta},
$$
\be
r'=e^{B_\beta}r,~~~~
F'=e^{-B^\top_\beta}F.
\ee
\vskip 3mm

$\bullet$ Transformations generated by $C_\beta$:
$$ A'=A,\quad\quad\quad r'=r,\quad\quad\quad F'=F,
$$
$$ B'=B+AC_\beta,
$$
\be
C'=C+(\dot C_\beta+C_\beta B+B^\top C_\beta)+C_\beta AC_\beta.
\ee
\vskip 3mm

$\bullet$ The diffeomorphism transformations:
\be
\lambda'_i(t)=\dot f(t)\lambda_i(f(t)),~~~~
r'(t)=r(f(t)),~~~~
F'(t)=F(f(t)).
\ee
\vskip 3mm

\section{Finite \W-transformations}

\hspace{\parindent}%
Now we will concentrate on the $Sp(4)$ model in order to obtain
\W-transformations associated to a partial gauge-fixing of the model.
The corresponding restricted model describes a coupling of the matter
variables $x_i$ to a world-line version of chiral \W-gravity.

The partial gauge-fixing we consider is
\be
\Lambda_r=\left(\begin{array}{cccc}
                H  &  0  &  0  &  1  \\
                0  & -H  &  1  &  0  \\
                C  &  T  & -H  &  0  \\
                T  &  F  &  0  &  H
              \end{array}\right).
\ee
In this gauge the action \bref{sp can act} becomes
\be
S=\int\dif t\left[(\dot x_1-H x_1)(\dot x_2+H x_2)+
\frac12\left(C x_1^2+2Tx_1x_2+Fx_2^2\right)
-F_1F_2\right].
\ee
The equations of motion arising from this action are:
\bea
\nonumber
&[L]_{x_1}=-(\frac\dif{\dif t}+H)(\dot x_2+Hx_2)+Cx_1+Tx_2,
\\\nonumber
&[L]_{x_2}=-(\frac\dif{\dif t}-H)(\dot x_1-Hx_1)+Fx_2+Tx_1,
\\\nonumber
&[L]_H=\dot x_1x_2-x_1\dot x_2-2Hx_1x_2,
\\\nonumber
&[L]_C=\frac12x^2_1,\quad
  [L]_F=\frac12x^2_2,\quad
  [L]_T=x_1x_2,
\eea
\be
[L]_{F_1}=-F_2,\quad\quad[L]_{F_2}=-F_1.
\ee

In order to find the gauge transformations of this action we can use the
infinitesimal or finite linear transformations of the $Sp(4)$ model.
We start by using the infinitesimal transformations found in the last section.
The matrix of gauge parameters is now parametrized as
\be
A_\beta=\left(\begin{array}{cc}
        \beta_2&\beta_{10}\\
        \beta_{10}&\beta_5
        \end{array}\right),\quad\quad
 B_\beta=\left(\begin{array}{cc}
        \beta_3&\beta_9\\
        \beta_8&\beta_6
        \end{array}\right),\quad\quad
 C_\beta=\left(\begin{array}{cc}
        \beta_1&\beta_7\\
        \beta_7&\beta_4
        \end{array}\right).
\ee

The residual transformations in the gauge (4.1) are parametrized
by $\beta_{10}$, $\beta_5$, $\beta_2$ and $\beta_6$.
The relations between the dependent and the independent parameters
are given by
\bea
\nonumber
&&\beta_1=-C\beta_{10}+\beta_5(-T+2H^2+\dot H)+2\dot\beta_5H
+\frac12\ddot\beta_5,
\\\nonumber
&&\beta_4=-F\beta_{10}+\beta_2(-T+2H^2-\dot H)-2\dot\beta_2H+\frac12
\ddot\beta_2,
\\\nonumber
&&\beta_7=-T\beta_{10}+\frac12\ddot\beta_{10}-\frac12\beta_2C-\frac12F
\beta_5,
\eea
\be
\beta_3=-\dot\beta_{10}-\beta_6,~~~~~
\beta_8=-\beta_5H-\frac12\dot\beta_5,~~~~~
\beta_9=\beta_2H-\frac12\dot\beta_2.
\ee

Let us consider the following redefinition of the gauge parameters:
\be
\nonumber
\eps=\beta_{10},~~~~
\\\nonumber
\alpha=\beta_6+H\beta_{10}+\frac12k\dot\beta_{10},
\ee
where $k$ is an arbitrary constant parameter.
The non-trivial expression of the parameter $\alpha$ is related to two facts:
1) the field $H$ can not be a primary field under reparametrizations
unless a term $H\beta_{10}$ is included,
2) The non-empty intersection between $\ker ad \bar{\Lambda}_r$ (where
$\bar{\Lambda}_r$ is equal to $\Lambda_r$ with all remnants gauge fields
put to zero) and the Cartan subalgebra of $Sp(4)$ algebra allows us to
introduce in a natural way the parameter $k$.

There are four infinitesimal remnant transformations corresponding to
the independent parameters $\epsilon, \alpha, \beta_2$ and $\beta_5$.
\vskip 3mm

$\bullet \eps$-sector.
\bea
\nonumber
&\hat{\delta} H=\eps\dot H+H\dot\eps+\frac12(k-1)\ddot\eps,
\\\nonumber
&\hat{\delta} T=\eps\dot T+2\dot\eps T-\frac12\dddot\eps,
\\\nonumber
&\hat{\delta} C=\eps\dot C+(3-k)C\dot\eps,
\\\nonumber
&\hat{\delta} F=\eps\dot F+(1+k)F\dot\eps,
\\\nonumber
&\hat{\delta} x_1=\eps(\dot x_1+F_2)+\frac12(k-2)x_1\dot\eps,
\\\nonumber
&\hat{\delta} x_2=\eps(\dot x_2+F_1)-\frac12kx_2\dot\eps,
\\\nonumber
&\hat{\delta} F_1=\eps(-\dot F_1-2HF_1+[L]_{x_1})-\frac12kF_1 \dot\eps,
\eea
\be
\hat{\delta} F_2=\eps(-\dot F_2+2HF_2+[L]_{x_2})-\frac12(2-k)F_2 \dot\eps.
\ee
Notice that the gauge fields transform as quasi-primary fields under
Diff; instead, the matter fields do not have nice transformation
properties under Diff. We can obtain the standard diffeomorphism
transformations for matter and auxiliary variables by introducing in
the corresponding transformation
an anti-symmetric combination of their equations of motion:

$$
\nonumber
\delta H = \hat{\delta} H, \quad \delta T = \hat{\delta} T,\quad
\delta C = \hat{\delta} C, \quad \delta F = \hat{\delta} F,
$$
\be
\\\nonumber
\delta q^i(t) = \hat{\delta}q^i(t) +
\int dt'\,M^{ij}(t,t')\,[L]_{q^j}(t'),
\\\nonumber
\ee
where
\be
q^1 = x_1,\quad q^2=x_2,\quad q^3=F_1,\quad q^4=F_2.
\ee
The only non-zero elements of $M^{ij}$ are:
$$
M^{13}(t,t')=M^{24}(t,t')=-M^{31}(t,t')=-M^{42}(t,t')=\eps(t)
\delta(t-t'),
$$
$$
M^{34}(t,t')=-\left(\dot\eps(t)+2\eps(t)H(t)\right)\delta(t-t') +
2\eps(t)\frac\dif{\dif t'}\delta(t'-t),
$$
\be
\\\nonumber
M^{43}(t,t')=-\left(\dot\eps(t)-2\eps(t)H(t)\right)\delta(t-t') +
2\eps(t)\frac\dif{\dif t'}\delta(t'-t).
\ee
 $M$ is an anti-symmetric matrix: $M^{ij}(t,t') = -M^{ji}(t',t)$.
Hence, $\delta$ can replace $\tilde\delta$ as the symmetry
transformation associated with $\eps$.
The matter and auxiliary variables transform as primary fields
under the new transformation,
\bea
\nonumber
&&\delta x_1 = \eps\dot x_1 + \frac12 (k-2) x_1\dot\eps,
\\\nonumber
&&\delta x_2 = \eps\dot x_2 - \frac12 k x_2\dot\eps,
\\\nonumber
&&\delta F_1 = \eps\dot F_1 - \frac12 (k-2) F_1\dot\eps,
\eea
\be
\delta F_2 = \eps\dot F_2 + \frac12 k F_2\dot\eps.
\ee
\vskip 4mm
$\bullet \alpha$-sector.
$$
\nonumber
\delta H=-\dot\alpha,\quad\quad\delta T=0,\quad\quad
\\\nonumber
\delta C=2\alpha C,\quad\quad\delta F=-2\alpha F,
$$
\be
\delta x_1=-\alpha x_1,\quad\quad\delta x_2=\alpha x_2,~~~~~~~~
\delta F_1=\alpha F_1,\quad\quad\delta F_2=-\alpha F_2.
\ee
\vskip 4mm

$\bullet \beta_2$-sector.
$$
\delta H=\frac12C\beta_2,\quad\delta T=\frac12\beta_2(\dot
C-2CH)+\dot\beta_2C, \quad\delta C=0,
$$
$$\delta F=\beta_2(4H^3-4HT-6H\dot H+\dot T+\ddot H)
$$
$$-\dot\beta_2(6H^2-2T-3 \dot H)+3H\ddot\beta_2-\frac12\dddot\beta_2,
$$
$$\delta x_1=\beta_2(2Hx_2+\dot x_2)-\frac12x_2\dot\beta_2+
\beta_2 F_1,\quad\delta x_2=0,
$$
\be
\delta F_1=0,~~~~~
\delta F_2=-\beta_2\left(\dot F_1-[L]_{x_1}\right)-\frac12\dot\beta_2F_1.
\ee

$\bullet \beta_5$-sector.
\bea
\nonumber
&&\delta H=-\frac12F\beta_5,\quad\delta T=\frac12\beta_5(\dot
F+2FH)+\dot\beta_5F, \quad\delta F=0,
\\\nonumber
&&\delta C=-\beta_5(4H^3-4HT+6H\dot H-\dot T+\ddot H)
\\\nonumber
&&\hspace{9mm}-\dot\beta_5(6H^2-2T+3\dot
H)-3H\ddot\beta_5-\frac12\dddot\beta_5,
\\\nonumber
&&\delta x_1=0,\quad\delta x_2=-\beta_5(2Hx_1-\dot
x_1)-\frac12x_1\dot\beta_5+\beta_5 F_2,
\eea
\be
\delta F_1=-\beta_5\left(\dot F_2-[L]_{x_2}\right)-\frac12\dot\beta_5 F_2,
{}~~~\delta F_2=0.
\ee

The algebra of these residual transformations is:

$$ [~\delta_\epsilon~,~\delta_{\epsilon'}~]~=~\delta_{{\epsilon''}};~~
{\epsilon}''~=~\epsilon'~\dot\epsilon~-\epsilon~\dot\epsilon',
$$
$$ [~\delta_\epsilon~,~\delta_{\alpha}~]~=~\delta_{{\alpha}'};~~
{\alpha}'=~-\epsilon~\dot\alpha
$$
$$ [~\delta_\epsilon~,~\delta_{\beta_2}~]~=~\delta_{{\beta_2}'};~~
{\beta_2}'=~(2-k)~\beta_2~\dot\epsilon-\epsilon\dot\beta_2,
$$
$$ [~\delta_\epsilon~,~\delta_{\beta_5}~]~=~\delta_{{\beta_5}'};~~
{\beta_5}'=~k~\beta_5~\dot\epsilon-\epsilon\dot\beta_5,
$$

$$ [~\delta_\alpha~,~\delta_{\beta_2}~]~=~\delta_{{\beta_2}'};~~
{\beta_2}'=~2~\alpha~\beta_2,
$$
$$ [~\delta_\alpha~,~\delta_{\beta_5}~]~=~\delta_{{\beta_5}'};~~
{\beta_5}'=~-2~\alpha~\beta_5,
$$
$$ [~\delta_\alpha~,~\delta_{\alpha'}~]~=
   [~\delta_\alpha~,~\delta_{\beta_{10}}~]~=0
$$
$$ [~\delta_{\beta_2}~,~\delta_{\beta_5}~]~=~\delta_{{\epsilon}'}+
\delta_{\alpha'}+\delta_{\beta_{10}'};
$$
$$
\epsilon'~=~{\beta_{10}'}~=~-{1\over 2}(\beta_2 \dot\beta_5-\dot\beta_2~
\beta_5)-2~H~\beta_2 \beta_5,
$$
$$\alpha'={1\over 4}(\dot\beta_2~\dot\beta_5-k~\beta_2~\ddot\beta_5
+(k-2)\beta_5~\ddot\beta_2)-(1+k)\beta_2~\dot\beta_5~H+
$$
$$
(3-k)~\beta_5 \dot\beta_2~H+\beta_2\beta_5(T-5H^2+(1-k)\dot H),
$$
\be
[~\delta_{\beta_2}~,~\delta_{{\beta_2}'}~]~=
   [~\delta_{\beta_5}~,~\delta_{{\beta_5}'}~]~=
   [~\delta_{\beta_2}~,~\delta_{\beta_{10}}~]~=
   [~\delta_{\beta_5}~,~\delta_{\beta_{10}}~]~=~0,
\ee
\vskip 3mm
where the $\beta_{10}$ transformation is a trivial transformation, i.e.
it is proportional to the equations of motion. It is explicitly given by
\bea
\nonumber
&&\delta H=~\delta T=~\delta C=~\delta F=0,
\\\nonumber
&&\delta x_1=\beta_{10}~F_2,~~~\delta x_2=\beta_{10}~F_1,
\\\nonumber
&&\delta F_1=-2\beta_{10}(\dot F_1+H~F_1)-\dot\beta_{10}~F_1-
\beta_{10}[L]_1,
\eea
\be
\delta F_2=-2\beta_{10}(\dot F_2-H~F_2)-\dot\beta_{10}~F_2-
\beta_{10}[L]_2.
\ee

Notice that  we have an open algebra with field-dependent structure
functions.
The non-closure of the present algebra is due to the introduction
of equations of motion in the definition of the $\eps$-transformation
in terms of the original $\beta$ transformations.

The matter and auxiliary variables $x_1$, $x_2$, $F_1$ and $F_2~$
transform as primary fields under diffeomorphisms with weights
$(\frac k2-1)$, $-\frac k2$, $(1-\frac k2)$ and $\frac k2$ respectively.
The gauge variables $C$ and $F$ transform also as primary fields with
weights $(3-k)$ and $(1+k)$.
Instead, $H$ and $T$ are quasi-primary fields with
weights $~1$ and $~2~$. The transformations with parameters
$\beta_2$, $\beta_5$ generate non-linear \W-diffeomorphisms. The
parameter $\alpha$ generates dilatations.

Now we are going to construct the finite form of the previous
transformations.
In order to do that we shall perform the partial gauge-fixing with the
help of the finite $Sp(4)$ transformations.
\vskip 3mm

$\bullet$ Diff. sector:~~
The residual finite diffeomorphisms are obtained by the following
composition of finite transformations:

$$
X\stackrel{\beta_7}{\longrightarrow} \Box
\stackrel{\beta_3,\beta_6}{\longrightarrow} \Box
\stackrel{{\rm diff}}{\longrightarrow}\tilde X,
$$
where $X$ stands for any variable. We can express
$\beta_7$, $\beta_3$, $\beta_6$ in terms of the Diff parameter $f$:
\be
\beta_7=\frac{{\ddot f}}{2{\dot f}},\quad~\beta_3={{k-2}\over
2}~\ln~\dot f,\quad~\beta_6=-{k \over 2}~\ln~\dot f.
\ee
and
$$
\left\{
\begin{array}{lll}
T & \rightarrow & {\dot f}^2\, T(f) - \frac{1}{2} \left( \frac{f^{(3)}}
{\dot f}- \frac{3}{2} \frac{{\ddot f}^2}{{\dot f}^2} \right)
\\
H & \rightarrow & {\dot f}\,H(f)
+ \frac12(k-1)\frac{\ddot f}{\dot f}
\\
C & \rightarrow & {\dot f}^{3-k}\,C(f)
\\
F & \rightarrow & {\dot f}^{1+k}\,F(f)
\end{array} \right.
$$ \\ $$
\rm{matter\ variables} \left\{
\begin{array}{lll}
x_1 & \rightarrow & {\dot f}^{\frac{k}{2}-1}\,x_1(f)
\\
x_2 & \rightarrow & {\dot f}^{-\frac{k}{2}}\,x_2(f)
\end{array} \right.
$$ \\
\be
\rm{auxiliary\ variables} \left\{
\begin{array}{lll}
F_1 & \rightarrow & {\dot f}^{1-\frac{k}{2}}\,F_1(f)
\\ \nonumber
F_2 & \rightarrow & {\dot f}^{\frac{k}{2}}\,F_2(f).
\end{array} \right.
\ee
\vskip 3mm

$\bullet \alpha$-sector (dilatations):~~
The finite transformation corresponding to the $\alpha$-sector
infinitesimal residual transformations is a composition of $\beta_3$ and
$\beta_6$ finite transformations:
$$
X\stackrel{\beta_3,\beta_6}{\longrightarrow}\tilde X,
$$
with
\be
-\beta_3~=~\beta_6~=~\alpha
\ee
and
$$
\left\{
\begin{array}{lll}
T & \rightarrow & T
\\
H & \rightarrow & H - {\dot \alpha}
\\
C & \rightarrow & e^{2\alpha}\, C
\\
F & \rightarrow & e^{-2\alpha}\, F
\end{array} \right.
$$ \\ $$
\rm{matter\ variables} \left\{
\begin{array}{lll}
x_1 & \rightarrow & e^{-\alpha}\,x_1
\\
x_2 & \rightarrow & e^{\alpha}\,x_2
\end{array} \right.
$$ \\
\be
\rm{auxiliary\ variables} \left\{
\begin{array}{lll}
F_1 & \rightarrow & e^{\alpha}\,F_1
\\
F_2 & \rightarrow & e^{-\alpha}\,F_2.
\end{array} \right.
\ee
\vskip 3mm

$\bullet \beta_2$-sector:~~
The residual finite $\beta_2$-diffeomorphisms are obtained by the
following composition of finite transformations:
$$
X\stackrel{\beta_4,\beta_7}{\longrightarrow} \Box
\stackrel{\beta_9}{\longrightarrow} \Box
\stackrel{\beta_2}{\longrightarrow}\tilde X,
$$
with
$$
\beta_9=~H \beta_2+\frac12 C \beta_2^2-\frac12 \dot\beta_2,~~~~
\beta_7=-\frac12 C \beta_2,
$$
\bea
\nonumber
&\beta_4=(2 H^2-T-\dot H)\beta_2
-2 H \dot \beta_2+\frac12 \ddot \beta_2+(2 H C-\frac12
\dot C)\beta_2^2&
\\
&-{3 \over 2} C \beta_2 \dot \beta_2+\frac12 C^2 \beta_2^3.&
\eea
and
$$
\left\{
\begin{array}{lll}
T & \rightarrow & T + \frac{1}{2}
\beta_2\,\dot C + C\,{\dot \beta_2} -
\beta_2\,C\,H - \frac{1}{4}{\beta_2}^2\,C^2
\\
C & \rightarrow & C
\\
H & \rightarrow & H + \frac{1}{2}\beta_2\,C
\\
F & \rightarrow & F + \beta_2\,\left(\dot T
- 6\,H\,\dot H + \ddot H - 4\,H\,T + 4\,H^3\right) +
\\
 & & {\dot \beta_2}\,\left(2\,T - 6\,H^2 + 3\,\dot H \right) +
3\,{\ddot \beta_2}\,H - \frac{1}{2}{\beta_2}^{(3)} +
\\
 & & {\beta_2}^2\,\left(5\,C\,H^2
- C\,T - 2\,C\,\dot H - 3\,\dot C \,H + \frac{1}{2}\ddot C \right) +
\\
 & & \beta_2\,{\dot \beta_2} \,\left(\frac{5}{2}\dot C - 8\,C\,H\right) +
\frac{7}{4}{{\dot \beta_2}}^2\,C +
\frac{3}{2} \beta_2\,{\ddot \beta_2}\,C +
\\
 & & {\beta_2}^3\left(2\,H\,C^2 - C\,\dot C \right) -
2\,{\beta_2}^2\,{\dot \beta_2} \,C^2 + \frac{1}{4}{\beta_2}^4\,C^3
\end{array} \right.
$$ \\ $$
\rm{matter\ variables} \left\{
\begin{array}{lll}
x_1 & \rightarrow & x_1 + \beta_2 \,\left({\dot x_2}+F_1+2\,H\,x_2\right)-
\frac{1}{2}{\dot \beta_2} \,x_2 + \frac{1}{2}{\beta_2}^2\,C\,x_2
\\
x_2 & \rightarrow & x_2
\end{array} \right.
$$
\be
\rm{auxiliary\ variables} \left\{
\begin{array}{lll}
F_1 & \rightarrow & F_1
\\
F_2 & \rightarrow & F_2 -
\beta_2\,({\dot F_1}-[L]_{x_1}) -
\frac{1}{2}\dot{\beta_2}\,F_1+\frac{1}{2}{\beta_2}^2\,C\,F_1.
\end{array} \right.
\ee
\vskip 3mm

$\bullet \beta_5$-sector:~~
The residual finite $\beta_5$-diffeomorphisms are obtained by the
following composition of finite transformations:
$$
X\stackrel{\beta_1,\beta_7}{\longrightarrow} \Box
\stackrel{\beta_8}{\longrightarrow} \Box
\stackrel{\beta_5}{\longrightarrow}\tilde X,
$$
with
$$
\beta_8=~-H \beta_5+\frac12 F \beta_5^2-\frac12 \dot\beta_5,~~~~
\beta_7=-\frac12 F \beta_5,
$$
\bea
\nonumber
&\beta_1=(2 H^2-T+\dot H)\beta_5
+2 H \dot \beta_5+\frac12 \ddot \beta_5
-(2 H F+\frac12 \dot F)\beta_5^2&
\\
&-{3 \over 2} F \beta_5 \dot \beta_5
+\frac12 F^2 \beta_5^3.&
\eea
and
$$
\left\{
\begin{array}{lll}
T & \rightarrow & T + \frac{1}{2}
\beta_5\,\dot F+F\,\dot{\beta_5}+\beta_5\,F\,H-\frac{1}{4}{\beta_5}^2\,F^2
\\
H & \rightarrow & H - \frac{1}{2}\beta_5\,F
\\
C & \rightarrow & C + \beta_5\,\left(\dot T-6\,H\,\dot H-\ddot H+
4\,H\,T-4\,H^3\right)
\\
 & & +\dot{\beta_5} \,\left(2\,T-6\,H^2-3\,\dot H\right)-
3\,\ddot{\beta_5}\,H-\frac{1}{2}{\beta_5}^{(3)}+
\\
 & & {\beta_5}^2\,\left(5\,F\,H^2-F\,T+2\,F\,\dot H+
3\,\dot F \,H+\frac{1}{2}\ddot F \right)+
\\
 & & \beta_5\,\dot{\beta_5}\,\left(\frac{5}{2}\dot F+8\,F\,H\right) +
\frac{7}{4}{\dot{\beta_5}}^2\,F + \frac{3}{2} \beta_5\,\ddot{\beta_5}\,F -
\\
 & & {\beta_5}^3\left(2\,H\,F^2+F\,\dot F \right) -
2\,{\beta_5}^2\,\dot{\beta_5}\,F^2 + \frac{1}{4}{\beta_5}^4\,F^3
\\
F & \rightarrow & F
\end{array} \right.
$$ \\ $$
\rm{matter\ variables} \left\{
\begin{array}{lll}
x_1 & \rightarrow & x_1
\\
x_2 & \rightarrow & x_2 + \beta_5\,\left(\dot{x_1}+F_2-2\,H\,x_1\right) -
\frac{1}{2}\dot{\beta_5}\,x_1 + \frac{1}{2}{\beta_5}^2\,F\,x_1
\end{array} \right.
$$
\be
\rm{auxiliary\ variables} \left\{
\begin{array}{lll}
F_1 & \rightarrow & F_1 -
\beta_5\,(\dot{F_2}-[L]_{x_2})-
\frac{1}{2}\dot{\beta_5}\,F_2+\frac{1}{2}{\beta_5}^2\,F\,F_2
\\
F_2 & \rightarrow & F_2.
\end{array} \right.
\ee

In this way we have integrated the non-linear \W-transformations
from the knowledge of the finite linear $Sp(4)$ transformations.
Notice the appearence of the schwarzian derivative in the transformation
of $T$.
We believe the procedure followed in this paper may be useful for
the study of finite
\W-transformations corresponding to general \W-algebras.
The possible difficulties of this approach arise in the election
of the independent parameters and gauge fields.

\section{Conclusions}

\hspace{\parindent}%
We have established a connection between zero-curvature condition of
$2d$ gauge theories and relativistic particle models with an $Sp(2M)$
gauge group. The zero-curvature condition
encodes a particular coupling between matter and \W-gravity.

We construct the finite linear gauge transformations of the $Sp(4)$
particle and perform a partial gauge-fixing at the level of finite
transformations.
We obtain its finite remnant gauge transformations. They can be
formally considered as classical chiral finite \W-transformations
of matter coupled to $Sp(4)$ \W-gravity.
A peculiarity of these transformations is the non-closure of the algebra
if we require the matter variables to transform as primary fields
under Diff.
A natural geometric interpretation of these transformations
can be given in the framework of the flag bundle approach to
zero-curvature condition \cite {G} \cite {BFK}.
The supersymmetric extension of this approach has been studied in
\cite {AG}.
Further details will appear elsewhere \cite {GHKR}.

\section*{Acknowledgements}

\hspace{\parindent}%
J. G. is grateful to Karyn Apfeldorf for discussions and to Prof.\
S. Weinberg
for the warm hospitality at the Theory Group of The University of Texas.
J. H. acknowledges a fellowship from the Generalitat de Catalunya and
J. G. is grateful to the Ministerio de Educaci\' on y Ciencia of Spain
for a grant.

This research was supported in part by Robert A. Welch Foundation,
NSF Grant PHY 9009850, NATO Collaborative
Research Grant (0763/87) and CICYT project no.\ AEN89-0347.

\end{document}